\documentclass[twocolumn,showpacs,preprintnumbers,amsmath,amssymb]{revtex4}

\usepackage{graphicx}
\usepackage{dcolumn}
\usepackage{bm}

\begin{document}

\preprint{CaAs-2003}

\title{Zinc-blende CaP, CaAs and CaSb as half-metals: \\
A new route to magnetism in calcium compounds
}

\author{Masaaki Geshi}
 \email{geshi@aquarius.mp.es.osaka-u.ac.jp}
\author{Koichi Kusakabe}
 \email{kabe@mp.es.osaka-u.ac.jp}
\author{Hidekazu Tsukamoto}
 \email{tsukamoto@aquarius.mp.es.osaka-u.ac.jp}
\author{Naoshi Suzuki}
 \email{suzuki@mp.es.osaka-u.ac.jp}
\affiliation{%
Graduate School of Engineering Science, 
Osaka University, \\
1-3 Machikaneyama-cho, Toyonaka, Osaka 560-8531, Japan
}%

\date{\today}

\begin{abstract}
Existence of ferromagnetism in bulk calcium compounds is discovered 
 theoretically. 
First-principles calculations of calcium phosphide, calcium arsenide and
 calcium antimonide in the zinc-blende structure have been performed to 
 show the half-metallic ground state in each optimized stable structure. 
Magnetism comes from spin-polarization of electrons in 
 $p$-orbitals of P, As or Sb and $d$-orbitals of calcium atoms. 
The half-metallicity is analogous to the half-metallic zinc-blende
 compounds, {\it e.g.} CrAs or CrSb, but the predicted compounds become 
 ferromagnetic without transition metals. 
In (In$_{1-x}$Ca$_x$)Sb, the magnetism remains to be stable in a
 range of the doping rate ($x>0.8$). 
\end{abstract}

\maketitle



In recent progress of spin electronics, discovery of novel magnetic 
 materials accelerates researches in this rapidly progressing
 field.\cite{Prinz}  
Almost complete search for possible magnetic materials has been 
 performed utilizing transition metal doping into III-V semiconductors or
 II-VI compounds.\cite{Sato1,Sato2} 
In addition to the dilute magnetic semiconductors (DMS),\cite{Matsukura} 
 new classes of materials were discovered as end materials of the doping. 
For the case of Mn-doping into GaAs, denser doping is not easy, but the
 theoretical calculation of the end material {\it i.e.} MnAs in the
 zinc-blende structure, shows that the material should be a 
 half-metal.\cite{Shirai0,Ogawa} 
For the case of Cr-doping into GaAs or InSb, end materials are known to
 possess the half-metallic band structures by the first-principles
 calculations.\cite{Shirai0,Shirai,Shirai2}  
Several ferromagnetic materials including zinc-blende CrAs (zb-CrAs) and
 zb-CrSb were indeed successfully synthesized based on the recent
 theoretical prediction.\cite{Akinaga,Zhao}

An interesting trend in the list of DMS obtained by doping of every 
 possible transition metal element is that the strong magnetism is realized
 when the $d$-shell is partially occupied by electrons and is not half-filled.  
Cr or Mn in GaAs and Co in ZnO seem to be optimal for the
 half-metallicity.\cite{Sato1,Sato2} 
Fe, Co, Ni are not necessarily suitable for the dopant. 
We would be able to obtain a better and deeper understanding of the
 magnetic effect in these curious materials by looking at all possible
 compounds including magnetic semiconductors. 

In this paper we report on first-principles electronic structure
 calculations of calcium compounds {\it i.e.} CaX (X=P, As and Sb). 
These materials are supposed to form the zinc-blende structure. 
In these crystals, the most interesting character of the band structure
 is that the highest occupied band becomes almost dispersionless. 
This curious flat band, which creates a sharp peak of the density of
 states (DOS) at the Fermi level, originates from hybridization of
 $p$-orbitals of pnictides and Ca 3$d$ orbitals.   
In this $p$-$d$ hybridized band, ferromagnetism arises. 
Indeed, our spin-dependent generalized-gradient-approximation (spin-GGA)
 calculation reveals that calcium phosphide, calcium arsenide and calcium
 antimonide in the zinc-blende structure (zb-CaP, zb-CaAs and zb-CaSb in
 short) possess half-metallic band structures. 

So far, several Ca compounds were discussed as 
 possible ferromagnetic materials. 
In CaB$_6$, the ferromagnetism reported in the literature was argued as
 a defect-mediated spin-polarization.\cite{Monnier}  
B$_6$ vacancies were supposed to be an origin of magnetism. 
Inter-orbital exchange interactions between degenerate molecular
 orbitals in a B$_6^{2-}$ cluster could be large enough to cause
 ferromagnetic behavior. 
Similar argument was proposed for CaO in the rock salt
 structure.\cite{Elfmov}  
The mechanism was derived from consideration of possible attraction
 between two holes in doubly degenerate molecular orbitals on a cluster
 of O$^{2-}$ surrounding a Ca vacancy. 
Since the exchange interaction between holes would be ferromagnetic, 
 there could be formation of a triplet state at the cluster, which was
 essential for the argument. 
Thus, in these examples, the magnetism was supposed to be favored due to
 degenerate orbitals on anion clusters but not on orbitals of Ca. 

In the case of zb-CaX, as shown in the following discussion, orbitals on
 calcium atoms play essential roles for the magnetism.  
Thus our finding of half-metallicity appearing in the zinc-blende Ca
 compounds is new. 
Here it might be meaningful to comment the superconductivity found in a 
 high pressure phase of Ca, since $d$-orbitals play 
 important roles.\cite{Okada} 
Existence of $d$-orbitals induced by the crystal field at Ca sites are
 not rare. 
Some theoretical studies of $d$-orbitals of Ca and/or alkaline earth
 elements and divalent rare earth elements, 
 are found in the literature.\cite{Duthie,Skriver} 

The first-principles band-structure calculations to determine the
electronic structure were done for each material by adopting the 
spin-GGA as the exchange-correlation energy functional.\cite{Perdew} 
The full-potential linearized augmented plane wave (FLAPW) method was
utilized. 
All of the numerical calculations with this method have been done using
the Wien2k code.\cite{Wien2k} 
$R_{MT} K_{max}$ value is fixed at 8.00, where $R_{MT}$ is a minimum
muffin-tin (MT) radius and $K_{max}$ is a maximum reciprocal lattice
vector.
MT radii are 0.22 $a$ for Ca and 0.18 $a$ for X, respectively, where $a$
is a lattice constant.
We have used an angular momentum expansion up to $l_{max}=10$.
The criterion of the energy convergence is set to 0.001 mRy and we
have as well checked the charge distance between the last two iterations
and that has converged less than 0.00001.
Five hundred k-points are taken in the first Brillouin zone.
Our calculational conditions are enough to determine the lattice
constant, the electronic structures and the magnetic moments,
accurately, except for overestimation of the lattice constants caused by
the spin-GGA calculation.
For InSb the present calculation overestimates the lattice constant by
2.6 \% as shown in Table \ref{table1}.
To investigate the case of Ca-doped compounds we have used a
Korringa-Kohn-Rostker method combined with a coherent potential
approximation (KKR-CPA)\cite{Akai} based on a local density
approximation.\cite{MJW}  
The form of the potential was restricted to the MT type.
The wavefuntions in the MT sphere were expanded in spherical waves with 
the angular momenta up to $l_{max}=2$. 
512 sampling $k$-points were taken in the first Brillouin zone.

\begin{table}
\begin{center}
\begin{tabular}{@{\hspace{\tabcolsep}\extracolsep{\fill}}rlllll} 
\cline{1-6}
                      &  CaP  & CaAs  & CaSb & InSb & InSb(Exp.)\\ 
Lattice constant[\AA] & 6.55  & 6.75  & 7.22 & 6.65 & 6.48\cite{InSb} \\
\cline{1-6}
\end{tabular}
\end{center}
\caption{The optimized lattice constants of zb-CaP, zb-CaAs and zb-CaSb
obtained by the spin-GGA calculation for the ferromagnetic states. 
The lattice constant of InSb determined by an experiment as well as an
 optimized value by the calculation is  shown for comparison.} 
\label{table1}
\end{table}

\begin{figure}
\includegraphics[width=8.5cm]{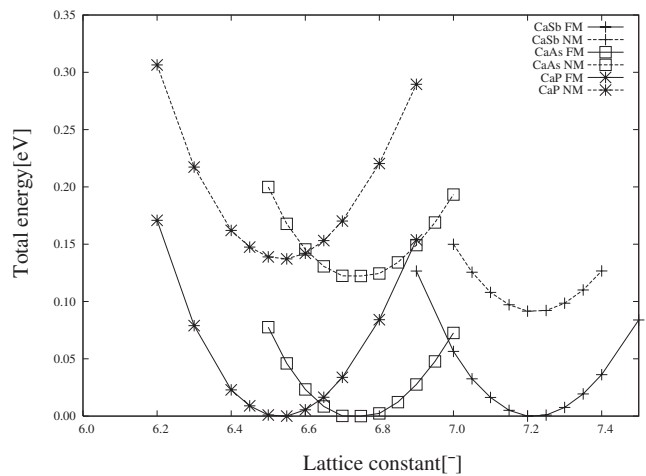}
\caption{\label{CaX-energy-volume} 
Energy-volume relation of zb-CaX (X=P,As,Sb). 
Each energy is shifted so that the stable ferromagnetic state has $E=0$
 for each compound. 
}
\end{figure}

The crystal structure of each zb-CaX was optimized in our calculation. 
The lattice constants were determined both for the ferromagnetic ground 
state and for the paramagnetic state.
Those are almost the same. 
Values for the ferromagnetic states are tabulated in Table
\ref{table1}. (See also Fig. \ref{CaX-energy-volume}.)
The crystal symmetry was unchanged in the optimization process 
irrespective of the magnetic state. 
In the zinc-blende structure, the ground state of zb-CaX becomes
spontaneously ferromagnetic. 
This conclusion was confirmed for the doubled unit cell which allows
zb-CaX to form the antiferromagnetic spin configuration. 
Thus we conclude that the ferromagnetic ground state appears at least in
the spin-GGA calculation.  
Energy gain by forming the ferromagnetic state from the paramagnetic 
state is about 0.09 eV for CaSb, 0.12 eV for CaAs and 0.14 eV for CaP. 
These values are comparable to those of other zinc-blende magnetic
compounds\cite{Shirai}.   

\begin{figure}
\includegraphics[width=8.5cm]{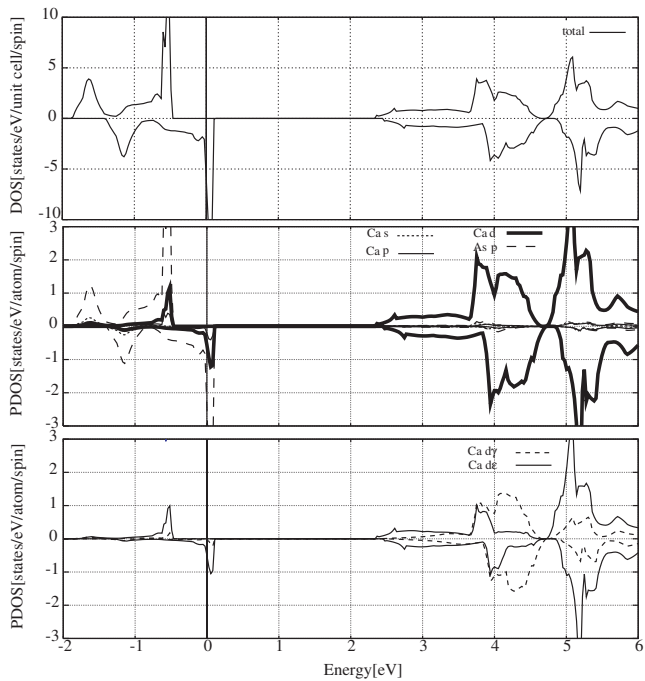}
\caption{\label{CaAs_dos} 
The DOS of zb-CaAs. 
The upper panel shows the total DOS, while the middle and lower
 represent the partial DOS decomposed into 4$p$-components of As, 3$s$-,
 3$p$- and 3$d$-components of Ca plotted in the broken, dotted, solid and
 bold solid lines in the middle panel, respectively,  and decomposed
 into $d\varepsilon$ and  $d\gamma$ states of Ca plotted in the broken
 dotted and solid lines in the lower panel, respectively. 
}
\end{figure}

\begin{figure}
\includegraphics[width=8.5cm]{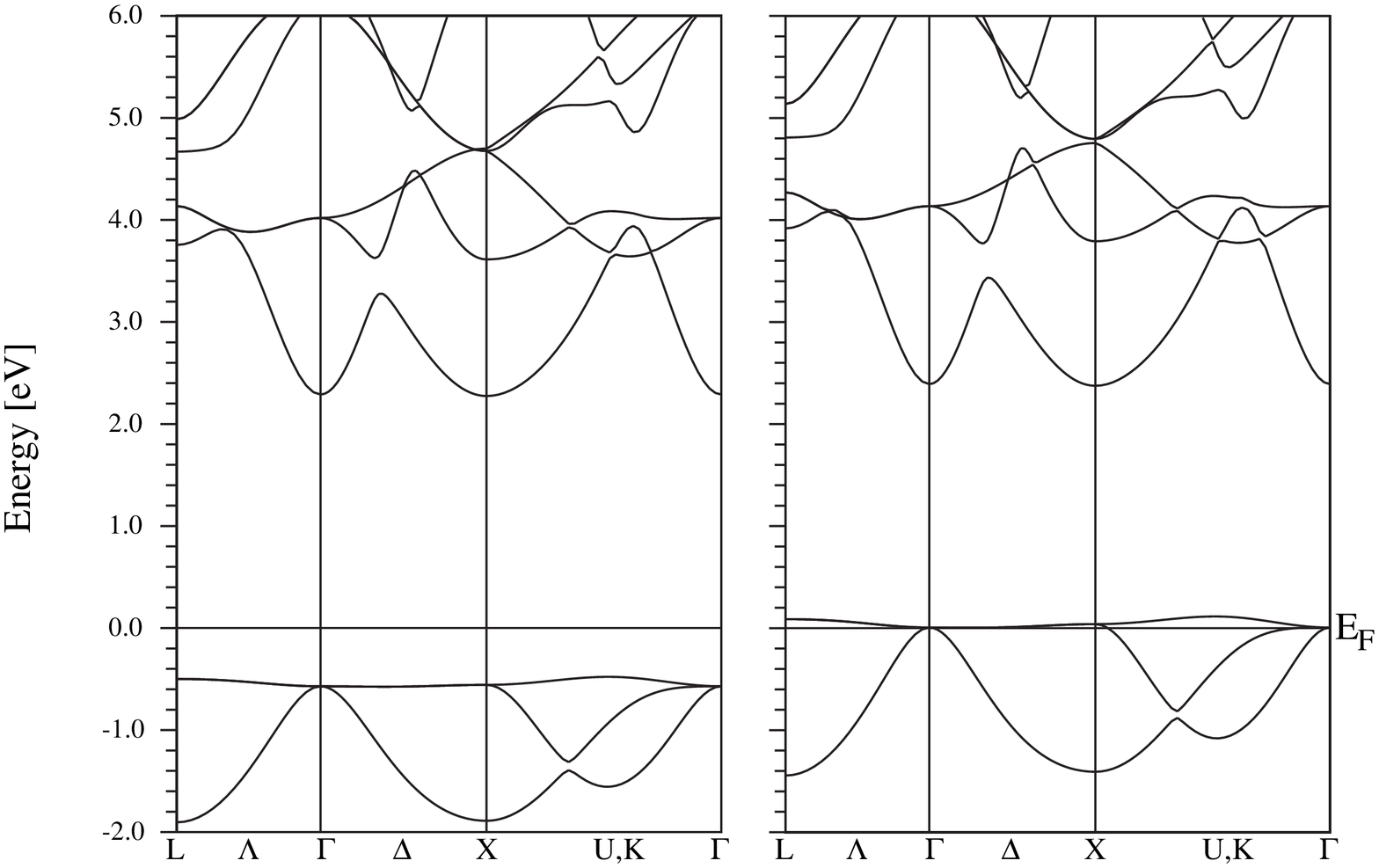}
\caption{\label{CaAs_band} 
The electronic band structure of the zb-CaAs. 
The left and right panels are spin-up and spin-down components, 
respectively. 
}
\end{figure}

We now discuss details of the electronic structure of zb-CaAs as a
typical example. 
The DOS of zb-CaAs is shown in Fig.\ref{CaAs_dos}. 
The valence band is strongly spin polarized with the exchange splitting
of $\sim$ 0.6eV.  
A sharp peak is found at the top of the valence band. 
The Fermi level locates at the peak of the minority-spin band. 
A gap opens in the majority-spin band and the present system is the 
half-metal.  
The origin of the curious peak of the DOS is an almost dispersionless band. 
(Fig.\ref{CaAs_band}) 
More precisely, a spin-down flat band forming the peak is empty,
while a spin-up flat band is filled by electrons. 
We can clearly see $d$-components of Ca in this dispersionless band. 
(See Fig.\ref{CaAs_dos}.) 
Although the valence band is almost composed of 4$p$-orbitals of As, the
dispersion of the top band is determined by the $p$-$d$ hybridization.
Note that $p$ bands become dispersive if As atoms are supposed to form
an FCC structure with the same lattice constant.  

To understand the magnetism, we specify characters of $d$-components of
Ca in the band structure of zb-CaX.  
Following the ligand field theory, we see that 
degenerate $d$-orbitals are split
into two levels in a crystal field with the tetrahedral symmetry.  
The $d\gamma$ level becomes the lower level and the $d\varepsilon$ level
becomes the upper one.  
The present result, of course, follows this rule and $d\gamma$ and
$d\varepsilon$ levels are at about 4.0 eV and 5.0 eV in the conduction
band, which is shown in the partial DOS (Fig. \ref{CaAs_dos}).  
Now the $d$-components appearing in the vicinity of the Fermi level
is $d\varepsilon$. (Fig. \ref{CaAs_dos}) 
This is because $d\varepsilon$ orbitals are strongly hybridized with
$p$-components of As. 
Formation of bonds is due to this $p$-$d$ hybridization in the
zinc-blede structure.  
This $p$-$d$ hybridization is often seen in magnetic compounds in the
zinc-blende structure.\cite{Sanvito,Zhao_F,Xu,Pask}  
Thus electrons prefer to occupy $d\varepsilon$ orbitals
rather than to do $s$-orbital or $d\gamma$.
The peak of the $d\varepsilon$ component in the vicinity of the Fermi level 
is not caused by the crystal field, but caused by the bond formation. 
These characteristics are common in zb-CaX. 

Because pnictide elements form a framework of the zinc-blende structure 
via $p$-$d$ hybridization, the valence band structure of zb-CaX is
basically similar to a series of compounds composed of 3$d$ transition
metals and V elements and/or VI elements. 
We now discuss similarity and difference between zb-CaX and other
compounds.   

The amount of the magnetic moment depends on occupation of electrons in
the $d$ levels. 
For the conventional transition metal compounds, for example, the
magnetic moment of MnAs is 4 $\mu_B$.  
Valence electrons are twelve (3$d^5$ 4$s^2$ from Mn and 4$s^2$
4$p^3$ from As) per a chemical formula.
However, two 4$s$ electrons of As are in quite deep energy levels 
at about -10 eV and these levels do not contribute to the bond formation and
magnetism. The rest ten electrons contribute to them.
Six of those electrons contribute to bond formation and four of them 
occupy $d$ levels and contribute to magnetism. 

With replacing Mn to Cr, V, Ti, and Sc, the magnetic moment per a 
transition metal atom decreases as 3, 2, 1, 0 $\mu_B$, respectively.
Because of Hund's rule coupling, 
spins of electrons occupying $d$ levels are parallel in each atom.
These high spin states are ferromagnetically aligned 
via the double exchange mechanism or by the $p$-$d$ exchange mechanism 
depending on the resulting electronic structure. 
This is the origin of magnetism of transition metal pnictides or
chalcogenides with the zinc-blede structure.

In zb-CaAs, five electrons (4$s^2$ of Ca and 4$p^3$ of As) contribute
to bond formation and magnetism.
Since one electron per a chemical formula is missing to fill the valence
$p$ bands, a hole band appears. 
Interestingly, at the top of the valence band, $d$ character is enhanced 
and the band becomes very flat. 
This fact suggests that interference effect occurs to make a localized
orbital per an As atom.  
Thus, localized nature appears at the top of the band.
The Fermi level comes just at this band and 
rather large exchange splitting appears 
due to the high DOS. 
As well as GaAs, the bonds of zb-CaAs are not strong covalent bonds and
they do not suppress magnetism. 

In zb-CaAs, however, there is no localized $d$ spins. 
Actually, polarized band is $p$ bands slightly hybridized with
$d$-orbitals of Ca. There is no room for the double exchange nor for the
$p$-$d$ exchange to work.  
The magnetism should be understood by another mechanism. 
The DOS of these curious $p$-$d$ hybridized bands looks like that of Ni
where the Fermi level comes at a peak position of the DOS. 
This is favorable for itinerant mechanism of ferromagnetic ordering. 
Here, it is worth commenting that the dispersionless band is just
half-filled. 
This situation is rather similar to the flat-band
ferromagnetism.\cite{Mielke}   


\begin{table}[t]
\begin{center}
\begin{tabular}{@{\hspace{\tabcolsep}\extracolsep{\fill}}rlll} 
\cline{1-4}
             &  CaP  & CaAs  & CaSb \\ 
Total        & 1.000 & 1.000  & 1.000 \\
Ca           & 0.092 & 0.110  & 0.139 \\
X            & 0.511 & 0.459  & 0.369 \\
Interstitial & 0.397 & 0.431  & 0.492  \\
\cline{1-4}
\end{tabular}
\end{center}
\caption{The magnetic moments of zb-CaP, zb-CaAs and zb-CaSb obtained by
 the spin-GGA calculation.
Those of Ca and X are defined in the MT spheres.
Interstitial represents an amount of the magnetic moment in 
the interstitial region.
} 
\label{table2}
\end{table}

The magnetic moments of zb-CaX are shown in Table \ref{table2}.
In all cases the total magnetic moment is 1 $\mu_B$.
Ca has only the tiny magnetic moment and P, As and Sb do have the
moments. 
A large amount of the moment is found in the interstitial region.  
The moments in the interstitial region are to be yielded mainly by $p$
of As slightly hybridized with $d$ of Ca, because most of the $d$ states
are in the MT spheres of Ca atoms. 
This result is consistent with the band structure as well as the DOS. 


\begin{figure}
\includegraphics[width=8.5cm]{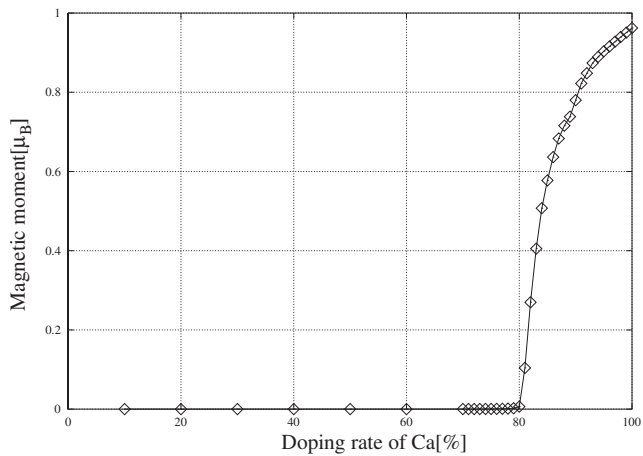}
\caption{\label{doping} 
The magnetic moment per a chemical formula as a function of Ca-doping
 rate in (In$_{1-x}$Ca$_x$)Sb. 
Ca atoms substitute In sites.
}
\end{figure}

We now investigate magnetism in Ca-doped compound, (In$_{1-x}$Ca$_x$)Sb.
The electronic states have been calculated by the KKR-CPA method.
In this calculation the experimental lattice constant (6.48 \AA) 
of InSb is used. 
As shown in Fig. \ref{doping} the magnetic moment appears at about 80 \%
of the doping rate and it increases monotonously to about 1 $\mu_B$.
We checked the DOS profile for each doping rate.
The sharp peak of the DOS originated from the dispersionless band
appears from low doping rate and it moves upward on the energy axis
with increasing doping rate. 
When the peak comes just at the Fermi level, the magnetic moment
appears. 


We have shown theoretically that Ca compounds in the zinc-blende
structure show ferromagnetism.  
The spin polarization occurs due to the exchange interaction in $p$-$d$
hybridized bands.  
The localized nature of the electronic states appears as a polarized
flat band at the top of the valence band.   
The half-metallic calcium compounds without the transition metals may
open new applications in spin electronics in addition to better
understanding of ferromagnetism found in several zinc-blende compounds.

This work was partly supported by the Ministry of Education, Culture,
Sports, Science and Technology of Japan (MEXT) under MEXT Special
Coordination Funds for Promoting Science and Technology (Nanospintronics
Design and Realization), a Grant-in-Aid for Scientific Research
(No.15GS0213), the 21st Century COE Program supported by Japan Society
for the Promotion of Science, NEDO under the Materials and
Nanogechnology Program and ACT-JST (Japan Science and Technology
Corporation) under Research and Development Applying Advanced
Computational Science and Technology program.
The calculation was partly done using the computer facility of ISSP,
University of Tokyo.

\bibliography{CaAs}

\begin{thebibliography}{25}
\expandafter\ifx\csname natexlab\endcsname\relax\def\natexlab#1{#1}\fi
\expandafter\ifx\csname bibnamefont\endcsname\relax
  \def\bibnamefont#1{#1}\fi
\expandafter\ifx\csname bibfnamefont\endcsname\relax
  \def\bibfnamefont#1{#1}\fi
\expandafter\ifx\csname citenamefont\endcsname\relax
  \def\citenamefont#1{#1}\fi
\expandafter\ifx\csname url\endcsname\relax
  \def\url#1{\texttt{#1}}\fi
\expandafter\ifx\csname urlprefix\endcsname\relax\def\urlprefix{URL }\fi
\providecommand{\bibinfo}[2]{#2}
\providecommand{\eprint}[2][]{\url{#2}}

\bibitem[{\citenamefont{Prinz}(1998)}]{Prinz}
\bibinfo{author}{\bibfnamefont{G.~A.} \bibnamefont{Prinz}},
  \bibinfo{journal}{Science} \textbf{\bibinfo{volume}{282}},
  \bibinfo{pages}{1660} (\bibinfo{year}{1998}).

\bibitem[{\citenamefont{Sato and Katayama-Yoshida}(2000)}]{Sato1}
\bibinfo{author}{\bibfnamefont{K.}~\bibnamefont{Sato}} \bibnamefont{and}
  \bibinfo{author}{\bibfnamefont{H.}~\bibnamefont{Katayama-Yoshida}},
  \bibinfo{journal}{Jpn. J. Appl. Phys.} \textbf{\bibinfo{volume}{39}},
  \bibinfo{pages}{L555} (\bibinfo{year}{2000}).

\bibitem[{\citenamefont{Sato and Katayama-Yoshida}(2001)}]{Sato2}
\bibinfo{author}{\bibfnamefont{K.}~\bibnamefont{Sato}} \bibnamefont{and}
  \bibinfo{author}{\bibfnamefont{H.}~\bibnamefont{Katayama-Yoshida}},
  \bibinfo{journal}{Jpn. J. Appl. Phys.} \textbf{\bibinfo{volume}{40}},
  \bibinfo{pages}{L334} (\bibinfo{year}{2001}).

\bibitem[{\citenamefont{Matsukura et~al.}(1998)\citenamefont{Matsukura, Ohno,
  SHen, and Sugawara}}]{Matsukura}
\bibinfo{author}{\bibfnamefont{F.}~\bibnamefont{Matsukura}},
  \bibinfo{author}{\bibfnamefont{H.}~\bibnamefont{Ohno}},
  \bibinfo{author}{\bibfnamefont{A.}~\bibnamefont{SHen}}, \bibnamefont{and}
  \bibinfo{author}{\bibfnamefont{Y.}~\bibnamefont{Sugawara}},
  \bibinfo{journal}{Phys.\ Rev.\ B} \textbf{\bibinfo{volume}{57}},
  \bibinfo{pages}{R2037} (\bibinfo{year}{1998}).

\bibitem[{\citenamefont{Shirai et~al.}(1998)\citenamefont{Shirai, Ogawa,
  Kitagawa, and Suzuki}}]{Shirai0}
\bibinfo{author}{\bibfnamefont{M.}~\bibnamefont{Shirai}},
  \bibinfo{author}{\bibfnamefont{T.}~\bibnamefont{Ogawa}},
  \bibinfo{author}{\bibfnamefont{I.}~\bibnamefont{Kitagawa}}, \bibnamefont{and}
  \bibinfo{author}{\bibfnamefont{N.}~\bibnamefont{Suzuki}},
  \bibinfo{journal}{J.\ Magn.\ Magn.\ Matter.}
  \textbf{\bibinfo{volume}{177-181}}, \bibinfo{pages}{1383}
  (\bibinfo{year}{1998}).

\bibitem[{\citenamefont{Ogawa et~al.}(1999)\citenamefont{Ogawa, Shirai, Suzuki,
  and Kitagawa}}]{Ogawa}
\bibinfo{author}{\bibfnamefont{T.}~\bibnamefont{Ogawa}},
  \bibinfo{author}{\bibfnamefont{M.}~\bibnamefont{Shirai}},
  \bibinfo{author}{\bibfnamefont{N.}~\bibnamefont{Suzuki}}, \bibnamefont{and}
  \bibinfo{author}{\bibfnamefont{I.}~\bibnamefont{Kitagawa}},
  \bibinfo{journal}{J.\ Magn.\ Magn.\ Matter.}
  \textbf{\bibinfo{volume}{196-197}}, \bibinfo{pages}{428}
  (\bibinfo{year}{1999}).

\bibitem[{\citenamefont{Shirai}(2001)}]{Shirai}
\bibinfo{author}{\bibfnamefont{M.}~\bibnamefont{Shirai}},
  \bibinfo{journal}{Physica\ E} \textbf{\bibinfo{volume}{10}},
  \bibinfo{pages}{143} (\bibinfo{year}{2001}).

\bibitem[{\citenamefont{Shirai}(2003)}]{Shirai2}
\bibinfo{author}{\bibfnamefont{M.}~\bibnamefont{Shirai}}, \bibinfo{journal}{J.\
  Appl. PHys.} \textbf{\bibinfo{volume}{93}}, \bibinfo{pages}{6844}
  (\bibinfo{year}{2003}).

\bibitem[{\citenamefont{Akinaga et~al.}(2000)\citenamefont{Akinaga, Manago, and
  Shirai}}]{Akinaga}
\bibinfo{author}{\bibfnamefont{H.}~\bibnamefont{Akinaga}},
  \bibinfo{author}{\bibfnamefont{T.}~\bibnamefont{Manago}}, \bibnamefont{and}
  \bibinfo{author}{\bibfnamefont{M.}~\bibnamefont{Shirai}},
  \bibinfo{journal}{Jpn.\ J.\ APpl.\ Phys.} \textbf{\bibinfo{volume}{39}},
  \bibinfo{pages}{L1118} (\bibinfo{year}{2000}).

\bibitem[{\citenamefont{Zhao et~al.}(2001{\natexlab{a}})\citenamefont{Zhao,
  Matsukura, Abe, Chiba, and Ohno}}]{Zhao}
\bibinfo{author}{\bibfnamefont{J.~H.} \bibnamefont{Zhao}},
  \bibinfo{author}{\bibfnamefont{F.}~\bibnamefont{Matsukura}},
  \bibinfo{author}{\bibfnamefont{E.}~\bibnamefont{Abe}},
  \bibinfo{author}{\bibfnamefont{D.}~\bibnamefont{Chiba}}, \bibnamefont{and}
  \bibinfo{author}{\bibfnamefont{H.}~\bibnamefont{Ohno}},
  \bibinfo{journal}{Appl. PHys. Lett.} \textbf{\bibinfo{volume}{79}},
  \bibinfo{pages}{2776} (\bibinfo{year}{2001}{\natexlab{a}}).

\bibitem[{\citenamefont{Monnier and Delly}(2001)}]{Monnier}
\bibinfo{author}{\bibfnamefont{R.}~\bibnamefont{Monnier}} \bibnamefont{and}
  \bibinfo{author}{\bibfnamefont{B.}~\bibnamefont{Delly}},
  \bibinfo{journal}{Phys. Rev. Lett.} \textbf{\bibinfo{volume}{87}},
  \bibinfo{pages}{157204} (\bibinfo{year}{2001}).

\bibitem[{\citenamefont{Elfmov et~al.}(2002)\citenamefont{Elfmov, Yunoki, and
  Sawatzky}}]{Elfmov}
\bibinfo{author}{\bibfnamefont{I.~S.} \bibnamefont{Elfmov}},
  \bibinfo{author}{\bibfnamefont{S.}~\bibnamefont{Yunoki}}, \bibnamefont{and}
  \bibinfo{author}{\bibfnamefont{G.~A.} \bibnamefont{Sawatzky}},
  \bibinfo{journal}{Phys. Rev. Lett.} \textbf{\bibinfo{volume}{89}},
  \bibinfo{pages}{216403} (\bibinfo{year}{2002}).

\bibitem[{\citenamefont{Okada et~al.}(1996)\citenamefont{Okada, Shimizu,
  Kobayashi, Ayama, and Endo}}]{Okada}
\bibinfo{author}{\bibfnamefont{S.}~\bibnamefont{Okada}},
  \bibinfo{author}{\bibfnamefont{K.}~\bibnamefont{Shimizu}},
  \bibinfo{author}{\bibfnamefont{T.}~\bibnamefont{Kobayashi}},
  \bibinfo{author}{\bibfnamefont{K.}~\bibnamefont{Ayama}}, \bibnamefont{and}
  \bibinfo{author}{\bibfnamefont{S.}~\bibnamefont{Endo}}, \bibinfo{journal}{J.
  Phys. Soc. Jpn.} \textbf{\bibinfo{volume}{65}}, \bibinfo{pages}{1924}
  (\bibinfo{year}{1996}).

\bibitem[{\citenamefont{Duthie and Pettifor}(1977)}]{Duthie}
\bibinfo{author}{\bibfnamefont{J.~C.} \bibnamefont{Duthie}} \bibnamefont{and}
  \bibinfo{author}{\bibfnamefont{D.~G.} \bibnamefont{Pettifor}},
  \bibinfo{journal}{Phys. Rev. Lett.} \textbf{\bibinfo{volume}{38}},
  \bibinfo{pages}{564} (\bibinfo{year}{1977}).

\bibitem[{\citenamefont{Skriver}(1982)}]{Skriver}
\bibinfo{author}{\bibfnamefont{H.~L.} \bibnamefont{Skriver}},
  \bibinfo{journal}{Phys. Rev. Lett.} \textbf{\bibinfo{volume}{49}},
  \bibinfo{pages}{1768} (\bibinfo{year}{1982}).

\bibitem[{\citenamefont{Perdew et~al.}(1996)\citenamefont{Perdew, Burke, and
  Ernzerhof}}]{Perdew}
\bibinfo{author}{\bibfnamefont{J.~P.} \bibnamefont{Perdew}},
  \bibinfo{author}{\bibfnamefont{S.}~\bibnamefont{Burke}}, \bibnamefont{and}
  \bibinfo{author}{\bibfnamefont{M.}~\bibnamefont{Ernzerhof}},
  \bibinfo{journal}{Phys. Rev. Lett.} \textbf{\bibinfo{volume}{77}},
  \bibinfo{pages}{3865} (\bibinfo{year}{1996}).

\bibitem[{\citenamefont{Blaha et~al.}(2001)\citenamefont{Blaha, Schwarz,
  Madsen, Kvasnicka, and Luitz}}]{Wien2k}
\bibinfo{author}{\bibfnamefont{P.}~\bibnamefont{Blaha}},
  \bibinfo{author}{\bibfnamefont{K.}~\bibnamefont{Schwarz}},
  \bibinfo{author}{\bibfnamefont{G.~K.~H.} \bibnamefont{Madsen}},
  \bibinfo{author}{\bibfnamefont{D.}~\bibnamefont{Kvasnicka}},
  \bibnamefont{and} \bibinfo{author}{\bibfnamefont{J.}~\bibnamefont{Luitz}},
  \bibinfo{journal}{WIEN2k, An Augmented Plane Wave + Local Orbitals Program
  for Calculating Crystal Properties (Karlheinz Schwarz, Techn.
  Universit\protect{\"{a}}t Wien, Austria)}  (\bibinfo{year}{2001}).

\bibitem[{\citenamefont{Akai}(1998)}]{Akai}
\bibinfo{author}{\bibfnamefont{H.}~\bibnamefont{Akai}}, \bibinfo{journal}{Phys.
  Rev. lett.} \textbf{\bibinfo{volume}{81}}, \bibinfo{pages}{3002}
  (\bibinfo{year}{1998}).

\bibitem[{\citenamefont{Morruzi et~al.}(1978)\citenamefont{Morruzi, Janak, and
  Williams}}]{MJW}
\bibinfo{author}{\bibfnamefont{V.~L.} \bibnamefont{Morruzi}},
  \bibinfo{author}{\bibfnamefont{J.~F.} \bibnamefont{Janak}}, \bibnamefont{and}
  \bibinfo{author}{\bibfnamefont{A.~R.} \bibnamefont{Williams}},
  \bibinfo{journal}{{\it Calculated Electronic properties of Metals}(Pergamon,
  New York, pp.11)}  (\bibinfo{year}{1978}).

\bibitem[{\citenamefont{Madelung}(1996)}]{InSb}
\bibinfo{author}{\bibfnamefont{O.}~\bibnamefont{Madelung}},
  \bibinfo{journal}{{\it Semiconductors - Basic data}, 2nd revised Edition
  (Springer-Verlag, Berlin Heidelberg New York pp.144)}
  (\bibinfo{year}{1996}).

\bibitem[{\citenamefont{Sanvito et~al.}(2001)\citenamefont{Sanvito, Ordej\'on,
  and Hill}}]{Sanvito}
\bibinfo{author}{\bibfnamefont{S.}~\bibnamefont{Sanvito}},
  \bibinfo{author}{\bibfnamefont{P.}~\bibnamefont{Ordej\'on}},
  \bibnamefont{and} \bibinfo{author}{\bibfnamefont{N.}~\bibnamefont{Hill}},
  \bibinfo{journal}{Phys. Rev. B} \textbf{\bibinfo{volume}{64}},
  \bibinfo{pages}{35207} (\bibinfo{year}{2001}).

\bibitem[{\citenamefont{Zhao et~al.}(2001{\natexlab{b}})\citenamefont{Zhao,
  Geng, Park, and Freeman}}]{Zhao_F}
\bibinfo{author}{\bibfnamefont{Y.~J.} \bibnamefont{Zhao}},
  \bibinfo{author}{\bibfnamefont{W.~T.} \bibnamefont{Geng}},
  \bibinfo{author}{\bibfnamefont{K.~T.} \bibnamefont{Park}}, \bibnamefont{and}
  \bibinfo{author}{\bibfnamefont{A.~J.} \bibnamefont{Freeman}},
  \bibinfo{journal}{Phys. Rev. B} \textbf{\bibinfo{volume}{64}},
  \bibinfo{pages}{35207} (\bibinfo{year}{2001}{\natexlab{b}}).

\bibitem[{\citenamefont{Xu et~al.}(2002)\citenamefont{Xu, Liu, and
  Pettifor}}]{Xu}
\bibinfo{author}{\bibfnamefont{Y.~Q.} \bibnamefont{Xu}},
  \bibinfo{author}{\bibfnamefont{B.~G.} \bibnamefont{Liu}}, \bibnamefont{and}
  \bibinfo{author}{\bibfnamefont{D.~G.} \bibnamefont{Pettifor}},
  \bibinfo{journal}{Phys. Rev. B} \textbf{\bibinfo{volume}{66}},
  \bibinfo{pages}{184435} (\bibinfo{year}{2002}).

\bibitem[{\citenamefont{Pask et~al.}(2003)\citenamefont{Pask, Yang, Fong,
  Pickett, and Dag}}]{Pask}
\bibinfo{author}{\bibfnamefont{J.~E.} \bibnamefont{Pask}},
  \bibinfo{author}{\bibfnamefont{L.~H.} \bibnamefont{Yang}},
  \bibinfo{author}{\bibfnamefont{C.~Y.} \bibnamefont{Fong}},
  \bibinfo{author}{\bibfnamefont{W.~E.} \bibnamefont{Pickett}},
  \bibnamefont{and} \bibinfo{author}{\bibfnamefont{S.}~\bibnamefont{Dag}},
  \bibinfo{journal}{Phys. Rev. B} \textbf{\bibinfo{volume}{67}},
  \bibinfo{pages}{224420} (\bibinfo{year}{2003}).

\bibitem[{\citenamefont{Mielke}(1993)}]{Mielke}
\bibinfo{author}{\bibfnamefont{A.}~\bibnamefont{Mielke}}, \bibinfo{journal}{J.
  Phys. Lett. A} \textbf{\bibinfo{volume}{174}}, \bibinfo{pages}{443}
  (\bibinfo{year}{1993}).

\end{thebibliography}

\end{document}